\documentclass[twocolumn,aps,pra,reprint,showpacs]{revtex4-1}

\usepackage{amsmath}
\usepackage{amsfonts}
\usepackage{amssymb}
\usepackage{hyperref}
\usepackage{graphicx}
\usepackage[font=large,caption=false]{subfig}
\usepackage[title,toc,titletoc]{appendix}
\usepackage{epstopdf}
\usepackage{mathtools}
\usepackage[mathscr]{euscript}
\usepackage[section]{placeins}
\begin{document}

\raggedbottom
\title{Continuous-variable quantum key distribution with a leakage from state preparation}
\author{Ivan Derkach}
\email{ivan.derkach@upol.cz}
\affiliation{Department of Optics, Palacky University, 17. listopadu 50,  772~07 Olomouc, Czech Republic}
\author{Vladyslav C. Usenko}
\email{usenko@optics.upol.cz}
\affiliation{Department of Optics, Palacky University, 17. listopadu 50,  772~07 Olomouc, Czech Republic}
\author{Radim Filip}
\email{filip@optics.upol.cz}
\affiliation{Department of Optics, Palacky University, 17. listopadu 50,  772~07 Olomouc, Czech Republic}
\date{\today}

\begin{abstract}
We address side-channel leakage in a trusted preparation station of continuous-variable quantum key distribution with coherent and squeezed states. We consider two different scenarios: multimode Gaussian modulation, directly accessible to an eavesdropper, or side-channel loss of the signal states prior to the modulation stage. We show the negative impact of excessive modulation on both the coherent- and squeezed-state protocols. The impact is more pronounced for squeezed-state protocols and may require optimization of squeezing in the case of noisy quantum channels. Further, we demonstrate that the coherent-state protocol is immune to side-channel signal state leakage prior to modulation, while the squeezed-state protocol is vulnerable to such attacks, becoming more sensitive to the noise in the channel. In the general case of noisy quantum channels the signal squeezing can be optimized to provide best performance of the protocol in the presence of side-channel leakage prior to modulation. Our results demonstrate that leakage from the trusted source in continuous-variable quantum key distribution should not be underestimated and squeezing optimization is needed to overcome coherent state protocols. 
\end{abstract}

\maketitle

\section{Introduction}

	Any practical realization of quantum key distribution (QKD) (see \cite{rev1, *rev2,*rev3} for reviews) deals with imperfections of real physical devices, which may be unaccounted in idealized security proofs. For example, it is well known that QKD systems based on direct photodetection [discrete-variable (DV) protocols] can be compromised by specific response of photodetectors to intense light, called blinding \cite{Lydersen2010}. On the other hand, an eavesdropper can implement so-called Trojan horse attacks in order to get information about the modulator settings from the back-reflected light \cite{trojan06, *tamaki16} or use state preparation and encoding flaws in DV QKD protocols \cite{Gott2002, *Tamaki, *Xu2014, Xu2010, *Wang16, *mizutani15} as well as benefit from information leakage, e.g., from auxiliary degrees of freedom of carrier states \cite{Nauerth}. Continuous-variable (CV) QKD protocols (see \cite{Braunstein2005, *Weedbrook2012, *Diamanti2015} for reviews), based on the homodyne detection, can be robust against blinding, but are potentially vulnerable to other practical attacks, such as a wavelength attack on the homodyne detector \cite{Huang2013} or continuous-variable counterpart of Trojan horse attacks \cite{Stiller}.\par

	 Most of the practical attacks on the QKD devices can be in principle ruled out using device-independent realization of QKD \cite{Vazirani2014} which, however, is very challenging (as it requires strongly entangled states and almost perfect detectors) and impractical, being limited to channels with high transmittance. There were also measurement-device independent (MDI) QKD protocols suggested and implemented, which rule out detector attacks \cite{Braunstein2012, *Lo2012}, but keep the source potentially vulnerable, while still being limited mostly to highly transmitting channels in the case of CV QKD \cite{Pirandola2015}.\par
Another method to make QKD more robust against practical imperfections and, at the same time, efficient and stable in conditions of strongly attenuating and noisy channels, is to distinguish between trusted devices (such as source and detector) and untrusted channel (the latter being under full control of an eavesdropper), which can be done by proper set-up characterization. Trusted parties can then identify possible sources of side information available to an eavesdropper, and take them into account in security analysis. In the field of CV QKD this included consideration of already mentioned specific detection attacks \cite{Huang2013, Ma2013} , analysis of source imperfections \cite{noise08, UF10, Jouguet2012, NoiseUF}, and role of multimode structure of state preparation and detection \cite{MultiMcv}. Trusted device imperfections may be under partial control of an eavesdropper so that an output of internal loss in a device may contribute to eavesdropper’s knowledge on the raw key though information leakage (side-channel loss) or so that the noise imposed by trusted device imperfections may be controlled by an eavesdropper to corrupt the data (side-channel noise). Such side channels, based on the basic linear coupling to vacuum or noisy modes, were previously considered on the detection and preparation sides of the protocol, assuming side-channel interaction after the modulation stage \cite{sidech1}. However, loss occurs as well on the stage of state preparation (e.g., it is well known that loss in the source reduces the level of squeezing \cite{Vahlbruch2016}). On the other hand, modulation can be applied to many modes at once \cite{MultiMcv} and some of the modes may be directly accessible by an eavesdropper which may result in a zero-error security break similar to a photon-number-splitting attack in DV QKD \cite{Huttner, *Brassard} enabled by multiphoton generation in a signal source. Therefore, in the current paper we analyze side-channel leakage in the trusted station prior to modulation (side-channel attack on the signal states) and also consider multimode modulation such that the auxiliary modes are directly available to an eavesdropper.  \par

	In our study we assume basic linear passive coupling with the side channels; we also assume that the trusted parties can be aware of the side channels presence in the trusted source (e.g., by characterizing their devices prior to and during the protocol implementation using local measurements; otherwise the side channel loss would be attributed to the main untrusted channel), but are not able to remove them and stop the potential information leakage. We consider two main classes of CV QKD protocols, namely coherent-state and squeezed-state Gaussian protocols. We show that both multimode modulation and side-channel attack on the signal can undermine security of CV QKD protocols. Moreover, such attacks appear to be surprisingly more harmful for the squeezed-state protocol, once the channel noise is low. For more noisy channels and combination of side-channel imperfections the protocol implementation should be optimized to provide security and maximum performance. \par
The paper is structured as follows. In Sec. \ref{Additional mode} we describe the mechanism of multimode modulation leakage, starting with the CV QKD model description and security analysis (Sec. \ref{Security}) following with the description of consequences for coherent- and squeezed-state protocols under individual and collective attacks \cite{Grosshans2003a, Grosshans2002, Navascues2006, Garcia-Patron2006} and distinction between direct and reverse reconciliation (Sec. \ref{Sq1}) \cite{Grosshans2003, cv3}. In Sec. \ref{SA} we first describe the model and methods used for security analysis of the side-channel attack on the signal states (Sec.\ref{SASecurity}) and further characterize the impact of such an attack on the security of CV QKD protocols with direct and reverse reconciliation under individual and collective attacks (Sec. \ref{SA part2}).
\section{Leakage from multimode modulator}
\label{Additional mode}

\subsection{Security analysis}
\label{Security}

\begin{figure*}
    \centering
    \hspace*{\fill}%
	\subfloat[Multimode modulation]{\label{fig:sub1}\includegraphics[width=.45\linewidth]{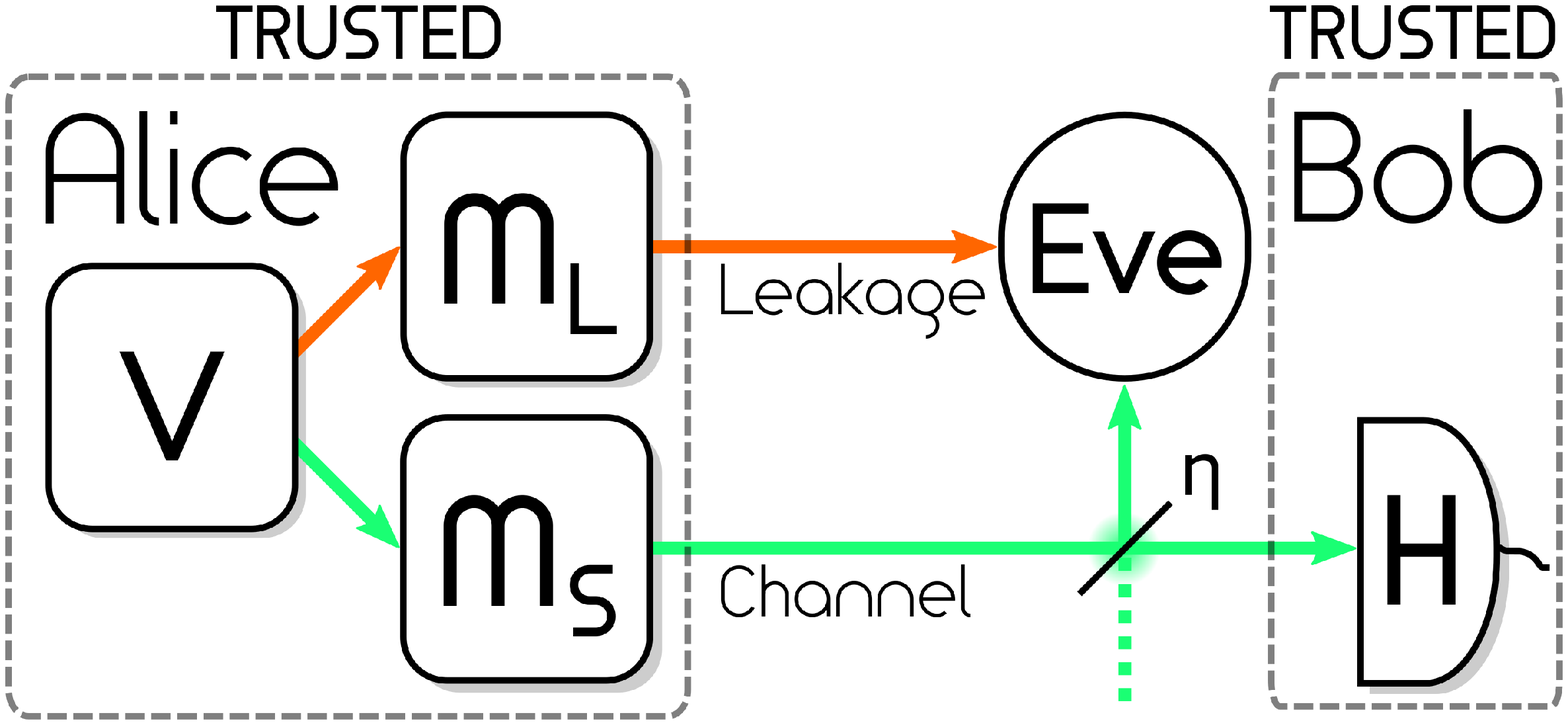}}
	\hfill 
	\subfloat[Premodulation channel]{\label{fig:sub2}\includegraphics[width=.45\linewidth]{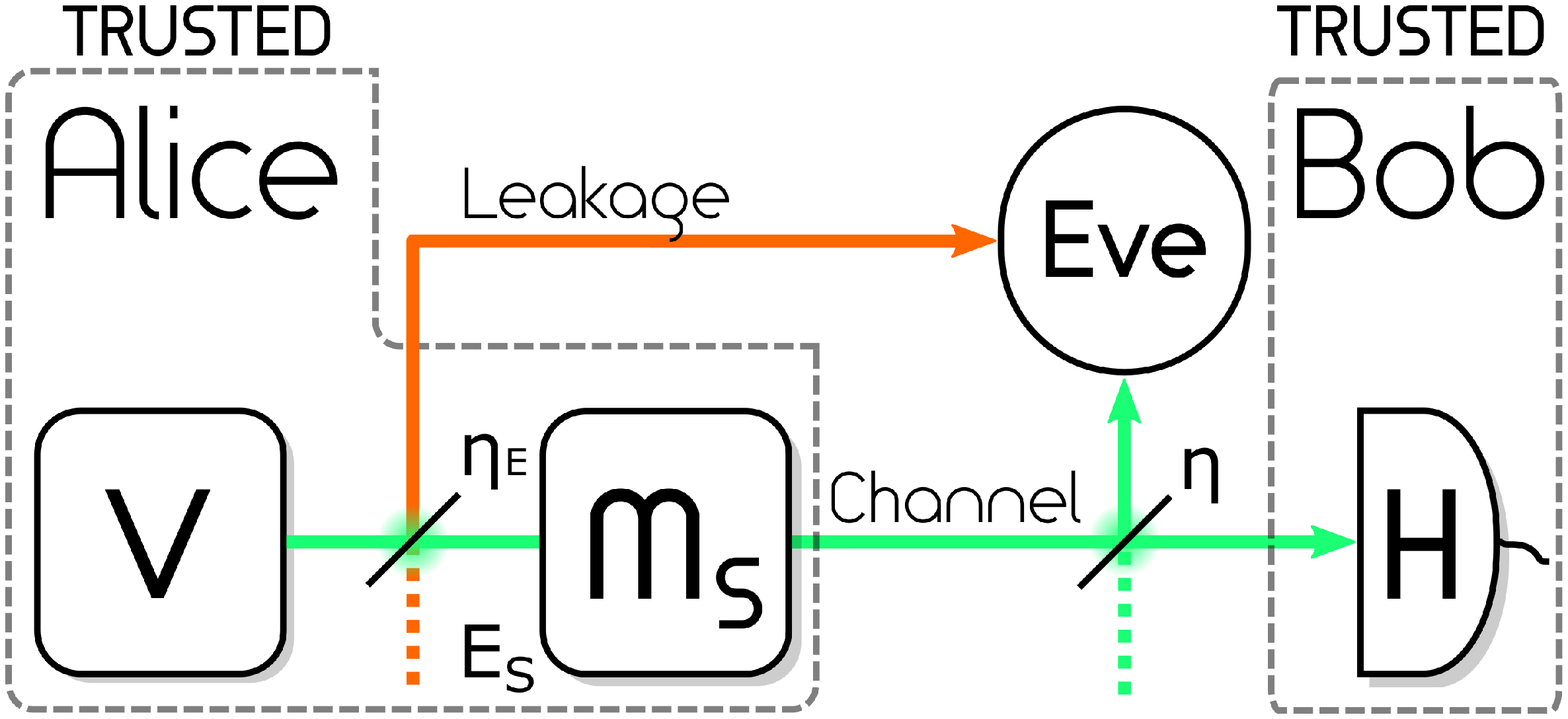}}
	\hspace*{\fill}
    \caption{Prepare-and-measure CV QKD schemes with lossy channels and information leakage from state preparation stations (dashed boxes indicate trusted stations of Alice and Bob). Source $V$ radiates Gaussian states (coherent or squeezed states) in the signal mode (green). States receive amplitude and phase displacements on modulator $M_S$, and are sent to Bob via quantum channel characterized by losses $\eta$. Signal states are measured on the receiver station by a homodyne detector $H$. \textbf{(a)} In addition to signal mode, the source generates additional leakage mode (orange line) $L$. States in the latter undergo displacement, correlated to the one of the main signal and characterized by modulation ratio $k$. An eavesdropper Eve can directly obtain information from additional mode as well as from quantum channel. The mode $L$ is present due to the multimode structure of the source and cannot be technically eliminated. Generally an arbitrary amount of modes $L_n$ can be modulated and leak, however such case can be reduced to a single effective mode $L_{eff}$.\textbf{(b)} A side-channel leakage (orange line) is present between state generation and state modulation stages. Initial signal state interacts with another state of mode $E_S$ on a beam-splitter with transmittance $\eta_E$, and only after that is being encoded with information on the modulator $M_S$. Eve can obtain information from $E_S$ and quantum channels.}
	\label{Scheme}
\end{figure*}

	We examine the effect of the presence and consequent modulation of signal states in additional modes generated by the source on the preparation side of a generic Gaussian CV QKD protocol, illustrated in Fig. \ref{fig:sub1}. Following the steps of a common CV QKD protocol \cite{Grosshans2003, Cerf2001, *Jouguet2013} the  trusted sender party prepares either coherent (using a laser source) or squeezed (using, e.g., the optical parametric oscillator) state characterized by the $X$ or $P$ quadrature (with both quadratures being interchangeable) value $Q_{S}$ with zero mean and variance $V_{S}=\langle Q_S^2 \rangle - \langle Q_S \rangle^2$ (for the coherent-state protocol $V_{S}=1$, while for the squeezed-state protocol signal quadrature variance $V_{S}<1$, so that the uncertainty relation is maintained as $V_{X}V_{P}\geqslant1$). Despite the state generated, Alice then applies both amplitude and phase quadrature modulation according to values $Q_M$ from two independent Gaussian distributions, with variance $V_M=\langle Q_M^2 \rangle - \langle Q_M \rangle^2$, to the output mode of the source so that the state entering the untrusted quantum channel and sent to Bob is characterized by the quadrature value $Q_B=Q_{S} +  Q_M$ and variance $V_B=V_{S} + V_{M}$.\par

	The source used by the sender can have a multimodal structure but it is usually presumed that Alice fully controls all the output of the source. In this work we assume that the source in addition to the main mode, characterized by the quadrature value $Q_{S}$ with variance  $V_{S}$, can produce additional $N$ leakage modes [Fig. \ref{fig:sub1}, orange line], which are characterized by the quadrature values $Q_{L_n}$ with respective variances $V_{L_n}$, that are not blocked or filtered by trusted parties. This results in amplitude and phase modulation being applied to the leakage modes as well. The signal state noise and modulation are trusted, but the leaking output is fully available to Eve.\\
Generally additional mode modulation $V_{M,L_n}$ may differ from the modulation $V_{M}$ applied to the signal mode, therefore we characterize the relation between them by the ratio $V_{M,L_n}/V_{M}=k$. If the $k=0$  additional mode is not modulated at all, this results in the state with the initial quadrature value $Q_{L_n}$, while for $k<1$ an additional mode receives a fraction of the signal modulation. Alternatively, leakage mode amplitude or phase quadrature displacements can correspond to a Gaussian distribution that has higher variance than that of the signal mode, corresponding to $k>1$. In other words the encoding alphabet of the secondary mode can be bigger than that of the signal mode, however excessive letters remain correlated to the signal alphabet. The signal state and the additional modulated state after the modulation are correlated as $C_{SL_n}=kV_{M}$, while leakage modes are correlated between each other as $C_{L_nL_m} = k^2 V_M$.\par

	After the preparation stage the signal $Q_B$ travels through the untrusted quantum channel (which is generally lossy and noisy, but for simplicity let us first consider the case of noiseless channel), where it is being measured by a homodyne detector. After the untrusted channel, Bob receives the state with quadrature values $Q_{B}'=(Q_{S} + Q_M)\sqrt{\eta} +Q_0 \sqrt{1-\eta} $ with variance $V_{B}'=(V_{S} +  V_{M}-1)\eta + 1$, where $Q_0$ is a quadrature value of the vacuum state that is coupled to the signal state in the channel and has variance $V_0=\langle Q_0^2 \rangle - \langle Q_0 \rangle^2=1$. An eavesdropper, after the signal passes through the untrusted channel, is able to acquire and store mode $E$ with $Q_{E}=-(Q_S +  Q_M)\sqrt{1-\eta} +Q_0 \sqrt{\eta} $ and variance $V_{E}=(V_{S} +  V_{M})(1-\eta) + \eta$, and additional source modes $L_n$ ($n \in [1,N]$) with $Q_{L_n}'=Q_{L_n}+kQ_{M}$ with variance $V_{L_n}'=V_{L_n}+k^2 V_{M}$.  After the signal state is transferred through the untrusted channel, initial correlations with the leakage mode are lowered by channel transmittance as $C_{SL_n}'=kV_M \sqrt{\eta}$.\par
To get analytical insights into the security of the protocol, and understand to basic limitations, we first study the case of individual attacks in a noiseless channel [as in Fig. \ref{fig:sub1}]. To purely see limitations by the leakage, we consider all data post-processing to be fully efficient. The lower bound on the secure key rate \cite{Devetak2005} under such attack is 

\begin{equation}
R|^{ind}_{RR(DR)}=I_{AB} - I_{BE(AE)},
\label{Rind}
\end{equation}
where $I_{XY}$ is the mutual information between respective parties, DR and RR stand for direct reconciliation (when Alice is the reference side of error correction) and reverse reconciliation (when Bob is the reference side \cite{Garcia-Patron2006}), respectively. The state measured by an eavesdropper, can consist of $(N+1)$ modes, including the untrusted quantum channel. Multimode modulation does not change the mutual information between trusted parties, and it corresponds to the one in conventional single-mode prepare-and-measure (P\&M) CV QKD protocols (binary logarithm indicates that units of information are bits) \cite{NoiseUF}:

\begin{equation}
I_{AB}=\frac{1}{2}\log_{2}\left[\frac{V_M}{V_M-\frac{\eta V^2_M}{\eta(V_S+V_M-1)+1}}\right].
\label{Iab1}
\end{equation}

Eve's mutual information with the trusted side depends on the variance of the state of a trusted party conditioned by the measurements of all the modes, available to Eve, $V_{A(B)|E}$ for direct or reverse reconciliation, respectively. For any $N$ leakage modes such a state can be reduced to $V_{A(B)|EL_{eff}}$, where $E$ is obtained from propagation losses in the quantum channel and $L_{eff}$ is the equivalent effective single-mode leakage. Second moments of the effective leakage mode in the signal quadrature and new effective modulation ratio can be, respectively, written as

\begin{equation}
V_{L_{eff}}=\frac{N}{\sum_{n}^{N}V_{L_n}^{-1}},
\label{Veff}
\end{equation}
\begin{equation}
k_{eff}=k\sqrt{N}.
\label{keff}
\end{equation}

In order to provide an extensive analysis of CV QKD protocols we examine the possible collective attacks that may be performed by Eve, resulting in the lower bound on the secure key rate given by
\begin{equation}
R|^{col}_{RR(DR)}=\beta I_{AB} - \chi_{BE(AE)},
\label{Key}
\end{equation}
where $\beta$ accounts for limited post-processing efficiency, mutual information $I_{AB}$ remains the same as in Eq. (\ref{Iab1}), while the information obtainable by the untrusted party is upper limited by the Holevo bound $\chi_{BE(AE)}$ \cite{Holevo2001} in either reverse or direct reconciliation. In the limit of an infinite block size Eq. (\ref{Key}) also corresponds to the key rate under coherent attacks \cite{CohAtt}.
Under collective attacks Eq. (\ref{Veff}) does not apply, however, second moments of the effective leakage mode can be found numerically. Nevertheless, provided that all $L_n$ have the same initial variance $V_{L}$, and multimode modulator [$M_L$ on Fig.\ref{fig:sub1}] outputs $N$ leakage modes $V_L'=V_L+k^2V_M$, the effective mode will have $V_{L,eff}=V_L$ with modulation ratio (\ref{keff}). We will further consider only the case with one additional mode ($L$) keeping in mind that a more general situation can be reduced to the single-mode one. The equivalent entanglement-based CV QKD scheme, enabling purification-based security analysis in the case of collective attacks \cite{NoiseUF} corresponding to Fig. \ref{fig:sub1}, due to the fact that a fraction of the correlated modulation leaked is unknown to trusted parties, is nontrivial. One way to find the solution is by applying the Bloch-Messiah reduction theorem \cite{Braunstein2005a} (for more details on security analysis see Appendix \ref{app1}). 

\subsection{Coherent- and squeezed-state protocols}
\label{Sq1}

\begin{figure}
\includegraphics[width=\linewidth]{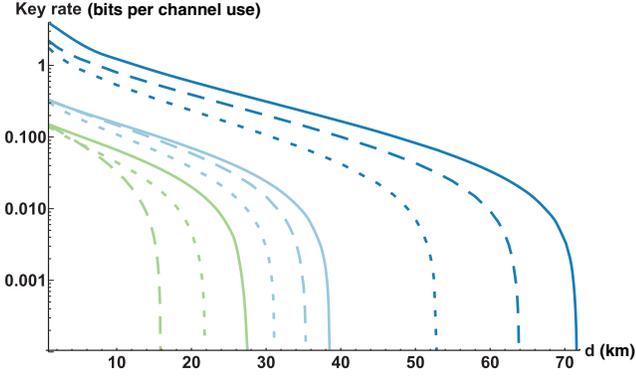}
\caption{Key rate (in bits per channel use) versus distance $d$ (in kilometers) in a standard telecom fiber (with attenuation of $-0.2dB/km$) under collective attacks in the case of modulation leakage for different values of ratio between additional and signal states modulation variances $k=0$ (blue, upper lines), 1 (light blue, middle lines), 1.5 (light green, lower lines) for optimized squeezed-state protocol (solid lines), squeezed-state protocol (dashed lines) with $V_{L}=V_{S}=1/2$ and coherent state protocol with $V_{L}=V_{S}=1$ (dotted lines). Modulation variance $V_M$ is optimized for given parameters, excess noise $\varepsilon=1\%$, post-processing efficiency $\beta=97\%$. Evidently the distance is shortened by modulation leakage. Squeezing optimization allows to achieve overall longer secure distances. Comparing unoptimized squeezing- and coherent-state protocols, the first one prevails under weak leakage $k\leqslant1$, the latter under stronger leakage $k>1$. }
\label{distance1}
\end{figure}

\textbf{Direct reconciliation.} This reconciliation scheme, which is more suitable for short distance channels, being limited by $-3dB$ of loss, is extremely sensitive to the information leakage from the additional source mode. In the limit of ideal state propagation through the quantum channel used by trusted parties $\eta \to 1$, and symmetry of the variances  $V=V_L=V_S$, the key rate (\ref{Rind}) reads
\small
\begin{equation}\label{DR0} 
   R_{DR} \approx \frac{1}{2}\bigg( \frac{[\eta-1]V_M}{V\log[2]}\frac{(2k^2V_M+V)^2}{k^2V_M+V}+\log_2 \bigg[\frac{V_M+V}{k^2V_M+V}\bigg]\bigg).
	\end{equation}
\normalsize
It is evident from Eq. (\ref{DR0}) that even if the quantum channel is perfect $(\eta=1)$ for arbitrary values of signal modulation the security is lost if the secondary mode receives the same modulation as the signal mode ($k=1$). In the absence of symmetry of variances $V_L\neq V_S$ excessive modulation can still lead to a security break even if input of the leaking modes are noisy coherent states with $V_L\geq 1$.

\textbf{Reverse reconciliation.} Again, assuming that all modes radiated by the source have the same variance $V=V_{L}=V_S$ in the limit of strong modulation ($V_{M} \to \infty$) the key rate (\ref{Rind}) reads

	\begin{multline}
R_{V_{M} \to \infty}|_{RR}^{ind} = - \frac{1}{2} \log_2 \bigg[ \left(1-\eta+ \frac{\eta k^2}{V(1+k^2)}\right)\\
\times (1+\eta[V-1])\bigg].
	\label{VtoInf}
	\end{multline}
If the leakage mode will be completely neglected trusted parties would underestimate Eves knowledge about the key that will lead to falsely estimated key rate:
	\begin{equation}
R^{(false)}_{V_{M} \to \infty}|_{RR}^{ind} = - \frac{1}{2} \log_2 \{ (1-\eta) [ 1+ \eta (V-1)] \}.
	\label{false}
	\end{equation}

	While mutual information (\ref{Iab1}) between Alice and Bob remains the same in Eqs. (\ref{VtoInf}) and (\ref{false}), the cost of underestimation of mutual information $V_{B|E}$ between Bob and Eve is $-1/2 \log_2 \{(1-\eta)/(1-\eta+k^2 \eta/[V(k^2+1)])\}$. Such cost for fixed $k$ is the highest for short distance $\eta \to1$ and high squeezing $V \to 0$, hence conditions which allow the high false key rate (\ref{false}) will in fact be security breaking and yield a negative actual key rate (\ref{VtoInf}).

	The correlated modulation $kV_M$ that leaks to the untrusted party makes the protocol sensitive not only to losses in the quantum channel $\eta$, but also to the initial state squeezing $V$ and the state modulation $V_M$; security is always limited by the presence of the second source mode for $\eta<1$. The more the squeezed initial state $V$ is, the smaller the fraction of the modulation $V_M$ is needed to be revealed to an eavesdropper to break the security of the protocol. In the limit of infinite squeezing $V \to 0$ for any nonzero modulation ratio $k$, the secure protocol cannot be established since the term contributing to Eve's information $k^2/ [V(1+k^2)]$ in Eq. (\ref{VtoInf}) approaches infinity, i.e., Eve is able to collect an accurate copy of the signal modulation directly from a leakage channel, without any attack on the main channel.\par
However, if the coherent-state protocol is used with $V=1$, one can see from Eq. (\ref{VtoInf}) that the secure key rate remains positive for any arbitrary amounts of correlated modulation leakage. For a long distance with small $\eta\ll 1$, we get always the positive secure key rate $\eta/(\ln 4(1+k^2))$. The key rate drops with longer distance, but never vanishes completely. \\
	Equation (\ref{VtoInf}) also allows one to assess the maximal tolerable $k_{max}$ ratio for high signal-state modulation:
\begin{equation}
k_{max}|_{V_M\to\infty}=\sqrt{\frac{V(\eta-2+V-\eta V)}{(\eta-1)(V-1)^2}},
\label{kmax}
\end{equation}
and immediately see that protocols can tolerate excess mode modulation with any ratio $k$ as long as either $\eta=1$ (quantum channel is perfect) or $V=1$ (coherent-state protocol is used). \\
Given that at $V=k^2/(1+k^2)$ the key rate (\ref{VtoInf}) becomes $R=-1/2 \log_2 [ 1-\eta+\eta k^2/(1+k^2)]$, and it is the same as when the coherent-state protocol is used ($V=1$), therefore the amount of squeezing needed to reach improvement over the coherent-state protocol is independent of channel losses $\eta$ and is bounded as

	\begin{equation}
\frac{k^2}{1+k^2}<V<1,
\label{range}
	\end{equation}

with squeezing that maximizes the key rate (\ref{VtoInf}) being

	\begin{equation}
V^{opt}|_{V_M\to \infty}=\sqrt{\frac{k^2}{1+k^2}}.
\label{vopt}
	\end{equation}

With the increase of the modulation ratio $k$ it is clear from Eqs. (\ref{range}) and (\ref{vopt}) that the coherent-state protocol is optimal in this regime, however, for low $k$, the optimized squeezed state protocol can yield significantly higher secure key rates.\par
One has to address an important aspect of the CV QKD system with multimode modulation---the difference between states in signal and leakage modes. Generally if the effective leaking state is initially more squeezed than the signal ($V_L<V_S$) it is more beneficial for an eavesdropper. An opposite effect is true as well---if the leaking state is initially less squeezed ($V_L>V_S$), the tolerance of protocols to modulation leakage is significantly improved, however, security is still limited by the leakage. For fixed state variance $V_L$ in the secondary source mode, optimal $V^{opt}_S<V_L$, provided $k<1$, but $V^{opt}_S>V_L$ if $k>1$.\par

	\begin{figure}
\includegraphics[width=.99\linewidth]{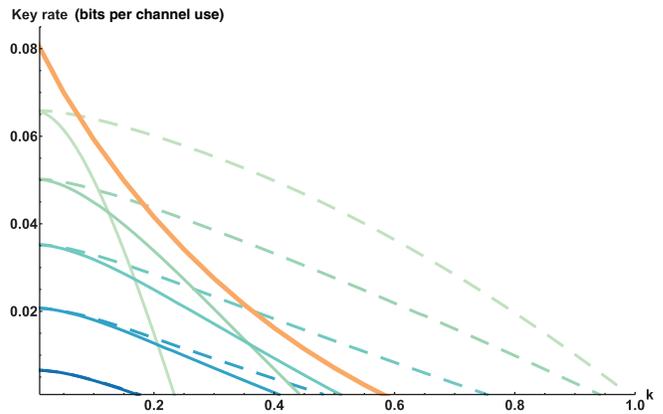}
	\caption{Performance of the squeezed-state protocol with and leakage from the modulator under collective attacks. The coherent-state protocol, due to combined effects of modulation leakage, excess noise and limited post-processing efficiency, cannot be used for secure key generation at given parameters. Key rate dependency on the leaked modulation ratio $k$ is shown for different values of squeezing (starting from top) $V_{S}$=0.1, 0.3, 0.5, 0.7, 0.9 SNU. All solid lines display the key rate with the symmetry of signal and leakage variances ($V_L=V_S$). The thick (orange) line illustrates the key rate of the protocol with both modulation $V_M$ and signal squeezing $V_S$ optimized. Dashed lines display the case when leakage mode input is fixed and independent of the signal ($V_L=1$). Lowest solid and dashed lines, corresponding to $V_S=0.9$ very nearly overlap. Signal modulation is optimized for given parameters. Reconciliation efficiency $\beta=95\%$,  channel losses $\eta=0.1$, and excess noise $\varepsilon=1\%$. Apparently the squeezed-state protocol is sensitive to leakage, especially for high squeezing. Provided that the source outputs identical states, security is broken when leakage is larger than half of the signal modulation. Robustness to leakage is higher if $V_L > V_S$, but leakage remains a security threat. Squeezing optimization (thick, orange line) heightens robustness and the key rate, but it's not sufficient to maintain security for arbitrary amounts of leakage.}

\label{allcoll3}
	\end{figure}
If noise is present in the channel one has to consider an equivalent entanglement-based CV QKD scheme for security analysis \cite{Grosshans2003b}. Results obtained for the protocols in realistic conditions (limited post-processing efficiency $\beta$ and noisy quantum channel, characterized by losses $\eta$ and noise $\varepsilon$) under collective attacks complement the preceding results for individual attacks. For any nonunity $\beta$, the signal modulation $V_M$ must be limited and optimized \cite{vmo1,*vmo2}. Leakage does have an impact on the optimal modulation value, but if the perfect post-processing algorithm $(\beta=1)$ is used, the key rate (\ref{Key}) as a the function of modulation $V_M$ is still monotonically increasing. Despite the states in the signal $V_S$ and leakage $V_L$ modes the excessive modulation is security breaking.\par
	Let us look at the case when the source generates identical states into signal and leakage modes $V=V_L=V_S$. In terms of secure distance (Fig. \ref{distance1}), the protocol with broadly accessible squeezing \cite{Andersen2015} of signal states to $-3 dB$ below the shot-noise unit (SNU) is able to prevail over the coherent-state protocol given the limited modulation ratio $k<1$. On the other hand the coherent-state protocol is less sensitive to leakage and can be used on longer distances if the modulation ratio is higher, $k>1$ (lower lines in Fig. \ref{distance1}). In fact, in such a regime even the noisy coherent-state protocol \cite{UF10, ThermalRW,Weedbrook2010} can achieve a higher key rate than squeezed-state protocol, provided excess noise $\varepsilon$ in the channel is low enough. \\
	However, in order to achieve best results under multimode modulation leakage with an arbitrary modulation ratio $k$, squeezing optimization is suggested. Optimal squeezing [similarly as in Eq. (\ref{vopt})] lowers with leakage increase, and approaches unity $V^{opt} \to 1$ for high $k$, e.g., in Fig. \ref{distance1} optimal squeezing $V^{opt}<1/2$ for $k=0, 1$ (upper and middle lines respectively), while for $k=1.5$ less squeezing is required $1>V^{opt}>1/2$ to achieve longer secure distance.\par
	Squeezed-state protocol susceptibility to leakage is further illustrated in Fig.\ref{allcoll3}. Highly squeezed states are clearly more sensitive to leakage, but lower squeezing does not necessarily yield higher tolerance to leakage. Contrary to the case of purely lossy channels, where the coherent-state protocol has a nonvanishing key rate for an arbitrary modulation leakage, it may not always be suitable for secure key generation in noisy channels. The squeezed-state protocol remains sensitive even if states in the leakage mode have fixed variance $V_L=const$ and are independent of signal $V_S$ (Fig. \ref{allcoll3}). Squeezing optimization in such a regime can still be effective, especially under strong leakage $k\gg1$.
\section{Premodulation leakage}
\label{SA}

\subsection{Security analysis}
\label{SASecurity}

In this section we will describe and examine another type of threat that may occur on a trusted preparation side of a Gaussian CV QKD system described in the beginning of the previous section (with absence of multimode modulation). Differently fom the previous section, we now consider the presence of a channel between the source and modulator, modeled as linear coupling to a vacuum mode, as shown in Fig.\ref{fig:sub2}. The signal generated by the source, with the quadrature value $Q_S$, prior to the modulation stage is linearly coupled to the mode $E_S$ with the coupling ratio $\eta_E$. Signal states have zero mean of quadratures and variances $V_{S}=\langle Q_S^2 \rangle - \langle Q_S \rangle^2$, while states in the premodulation channel $E_S$ are vacuum. During modulation (on modulator $M_S$) Alice applies displacement $Q_M$ with $\langle Q_M^2 \rangle - \langle Q_M \rangle^2=V_M$ to both quadratures of the signal states, resulting in a state $Q_B=Q_S\sqrt{\eta_E}+Q_{E_S}\sqrt{1-\eta_E}+Q_M$ with variance $V_B=\eta_E (V_{S}-1)+V_{M}+1$. Eve can gain information from states $Q_{E_S}'=Q_{E_S}\sqrt{\eta_E}- Q_S\sqrt{1-\eta_E}$ and $Q_{E}'=Q_E\sqrt{\eta}- \left(Q_S\sqrt{\eta_E}+Q_{E_S}\sqrt{1-\eta_E}+Q_M\right)\sqrt{1-\eta}$, obtained respectively from quantum and premodulation channels. After the signal passes through the purely lossy untrusted channel and arrives at the trusted receiver side its variance before measurement is

\begin{equation}
V_{B}'= [\eta_E (V_{X(P),S}-1)+V_{M}]\eta+1,  
\label{SA_VES}
\end{equation}
where $\eta$ characterizes the loss rate in the transmitting channel. General correlations between Eve's states are described as
\begin{equation}
C_{E_S,E_C}=(V_{S}-V_{E_S})\sqrt{(1-\eta)(1-\eta_E) \eta_E},
\label{CEE}
\end{equation}
while correlations between the signal state and output of the premodulation channel are scaled by the transmittance in the quantum channel,
\begin{equation}
C_{B,E_S}=(V_{E_S}-V_{S})\sqrt{\eta(1-\eta_E)\eta_E}.
\label{CBE}
\end{equation}
Using the expressions above we can write the mutual information between Alice and Bob as
\begin{equation}
I_{AB}=\frac{1}{2}\log_2{\frac{V_M}{V_M-\frac{V^2_M \eta}{V_M\eta+(V_S-1)\eta_E\eta+1}}}.
\label{Iab2}
\end{equation}
Similarly we can find the mutual information between Eve and the respective trusted reference side and apply Eq. (\ref{Rind}) for analysis of individual attacks. For general analysis we adopt the recently introduced general purification scheme \cite{sidech1} and proceed with an estimation of the CV QKD protocols behavior under collective attacks Eq. (\ref{Key}) in noisy quantum channels. 

\subsection{Coherent- and squeezed-state protocols}	
\label{SA part2}

The main aspect of the premodulation channel is that it provides correlations (\ref{CBE}) with the signal to the external party and corrupts the initial carriers states (\ref{SA_VES}). The influence of such a channel can be viewed as a preparation noise \cite{NoiseUF, UF10}, however, it also provides an eavesdropper with additional correlations with the signal. Furthermore, the premodulation channel can be equivalent to side-channel leakage after the modulation stage, provided $Q_M$ is scaled by $\sqrt{\eta_E}$ \cite{sidech1}.\\
For $V_S=1$, assuming the initial state in the mode $E_S$  is a vacuum state ($V_{E_S}=1$) we can immediately conclude that such a lossy channel would not affect the coherent-state protocol, since correlations (\ref{CEE}) and (\ref{CBE}) will totally vanish, and mutual information between trusted parties (\ref{Iab2}) will turn to a conventional form (\ref{Iab1}). In other words, the access to the lossy side channel would not provide the eavesdropper with any additional advantage if the coherent-state protocol is used. For $V_S<1$ and $V_{E_S}=1$ the correlation arises and this case has to be analyzed in detail. \\

\begin{figure}
\includegraphics[width=.95\linewidth]{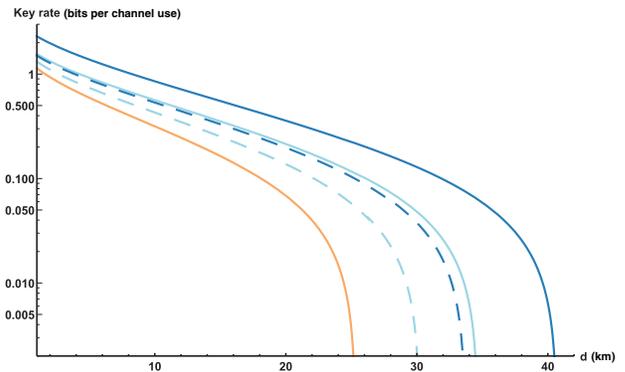}
	\caption{The key rate (in bits per channel use) versus distance $d$ (in kilometers) in a standard telecom fiber (with attenuation of $-0.2dB/km$) in the case of collective attacks on the coherent-state protocol (orange, lower line) and the squeezed-state protocol with $V_S=1/10, 1/2$ (upper, dark blue and middle, light blue respectively). The premodulation channel coupling ratio $\eta_E=0.5$ (dashed lines) and $1$ i.e., the absence of the channel (solid lines). Modulation variance is optimized for given parameters, $\beta = 97\%$, $\epsilon=5\%$. Evidently the premodulation channel reduces the secure distance of the squeezed-state protocol. However, even small squeezing allows one to achieve longer distances. Maximal influence of the premodulation channel is set by the performance of the coherent-state protocol.}
\label{distance}
\end{figure}

\textbf{Direct reconciliation.} Squeezed-state protocol is affected by the presence of the side channel between the source and modulator, since mutual information (\ref{Iab2}) diminishes, while information obtained by Eve (expressed in terms of mutual information or Holevo bound for respective attacks) increases. The lower bound on the key rate (\ref{Rind}) in a perfectly transmitting channel can be expressed as 

\begin{equation}
R|^{ind}_{DR}|_{\eta \to 1} = \frac{1}{2} \log_2 \bigg[ 1+ \frac{V_M}{1+ \eta_E (V_S-1)}\bigg].
\end{equation}

The presence of the premodulation channel lowers the overall key rate, however, secure key distribution is still possible for any side channel coupling ratio $\eta_E$. 

\textbf{Reverse reconciliation.} Similarly the squeezed-state protocol key rate (\ref{Rind}) will decrease due to the existence of the premodulation channel:

\begin{equation}
R|^{ind}_{RR}|_{V_M \to \infty}=-\frac{1}{2} \log_2 \big[ (1- \eta) (1+\eta_E (V_S-1)\eta ) \big],
\end{equation}

 though the key rate will still exceed the one for the coherent-state protocol. In the case of individual attacks and in the limit of high modulation $(V_M\to \infty)$ the advantage of the squeezed-state over the coherent-state protocol is

\begin{equation}
	\left(R|^{sq}_{RR}-R|^{coh}_{RR}\right)^{ind}_{V_M\to\infty}=-\frac{1}{2}\log_2 [1+\eta_E (V_S-1) \eta].
	\label{SoverC}
\end{equation} 

The premodulation channel grants adversary correlations with the signal and they provide Eve an additional advantage, comparing to the case of preparation noise \cite{UF10}. Such an advantage diminishes for low transmittance quantum channels and is the highest for $\eta\to1$:

\begin{equation}
	 R_{E_S} -R_{\Delta V}=\frac{1}{2}\log_2\left[ \frac{1+V_M+\eta_E(V_S-1)}{V_M+V_S/(\eta_E+V_S-\eta_EV_S)} \right].
	\label{advantage}
\end{equation} 

The correlations advantage $ R_{E_S} -R_{\Delta V}$ quickly disappears for the high modulation $(V_M\to\infty)$. \par 
Considering the noisy quantum channel and equivalent entanglement-based system under collective attacks  (see Appendix \ref{app2}) premodulation channel impact is similar to the previously described case of multimode modulation. the squeezed-state protocol is still superior to the coherent-state one in terms of the secure key rate (\ref{Key}) and tolerance to the channel noise. Squeezing optimization is not required as the key rate (\ref{Key}) linearly increases with an increase of squeezing. While the premodulation channel does not pose a security breaking threat, it can lower the secure distance (Fig. \ref{distance}) and tolerance to the quantum channel excess noise $\varepsilon$. Even though correlations (\ref{CBE}) help an adversary, the worst case scenario for trusted parties is substitution of the initial squeezed state by the coherent states ($V_S=1$).\\

\section{Conclusions}
We have investigated the negative impact of leakage from trusted the preparation side, namely the correlated multimode modulation of non-signal modes of the source and signal loss prior to the modulation stage. We have considered CV QKD coherent- and squeezed-state protocols with direct and reverse reconciliation. We have analyzed prepare-and-measure and equivalent entanglement-based models of leakage for cases of an illustrative individual and more general collective attacks in noisy channels. \par
Multimode modulation of nonsignal modes of the source limits the performance of both protocols and can lead to a security break even in the case of individual attacks in a purely lossy channel. 
Surprisingly, the coherent-state protocol can tolerate arbitrary amounts of leakage, though only in the noiseless channel. On the other hand, security of the squeezed-state protocol, with an increase of modulation leakage, quickly becomes compromised without the need for an untrusted party to resort to any additional manipulations onto the trusted side. We show that squeezing, however, can be optimized in order to improve the tolerance against multimode modulation leakage and channel noise. The optimized squeezed protocol then overcomes the coherent state protocol for any parameters. \par
The leakage from the preparation side prior to the modulation stage introduces noise to the squeezed signal and establishes correlations with an eavesdropper. While the coherent-state protocol is immune to such influence, the squeezed-state protocol suffers from secure key rate deterioration and becomes more sensitive to the excess noise in the channel. Nevertheless performance of the squeezed-state protocol surpasses the one of the coherent-state protocol, without the need for squeezing optimization.\\ 
Our results together with previous studies \cite{noise08,sidech1} describe the effects of the main possible mechanisms of information leakage from the trusted preparation side of continuous-variable quantum key distribution protocols, based on the most common linear passive coupling between optical modes. The results are stimulating for an experimental test of the macroscopically multimode protocols \cite{Usenko2015, Iskhakov2009, *Iskhakov2016, Christ2012, *Harder2014, Pinel2012, *Roslund2014}. They may also stimulate analysis of side channels in MDI CV QKD protocols, where side-channel attacks on the source are, in principle, possible and therefore relevant similarly to the case of discrete-variable protocols \cite{attack1}.

\noindent \textbf{Acknowledgement} The authors acknowledge support from Project No. LTC17086 of the INTER-EXCELLENCE program of Czech Ministry of Education. I.D. acknowledges funding from Palacky University Project No.the IGA-PrF-2017-008. 

\newpage
	\onecolumngrid
	\appendix
\section{LEAKAGE FROM THE MULTIMODE MODULATOR} \label{app1}

\subsection{Multimode leakage}

Using the initial values of quadrature variances of signal, leakage and untrusted channel modes (described in the main text), and input-output relations [for arbitrary modes 1 and 2 with quadrature vectors $\upsilon_i=(x_i,p_i)^T$],

\begin{equation}
	\label{inout}
	\left(\begin{matrix} \upsilon_1 \\ \upsilon_2 \end{matrix}\right)_{out}=\left( \begin{matrix} \sqrt{T}\mathbb{I} & \sqrt{1-T}\mathbb{I} \\ -\sqrt{1-T}\mathbb{I} & \sqrt{T}\mathbb{I}\end{matrix}\right) \left(\begin{matrix} \upsilon_1 \\ \upsilon_2 \end{matrix}\right)_{in},
\end{equation}

 one can obtain the results of linear interactions in prepare-and-measure (P\&M) multimode modulation scheme.  Provided the source radiates modes with identical variance ($V_L=V_S$) and the untrusted channel is purely lossy, 

\begin{equation}
\label{ble}
\sigma_{B,LE}=\left(
\begin{array}{cc}
 \sqrt{\eta} k V_M & 0 \\
 0 & -\sqrt{\eta} k V_M \\
 -\sqrt{(1-\eta) \eta} (V_M+V_S-1) & 0 \\
 0 & \frac{\sqrt{(1-\eta) \eta} (V_S-V_M V_S-1)}{V_S} \\
\end{array}
\right),
\end{equation}

\begin{equation}
\gamma_{LE}=\left(
\begin{array}{cccc}
 V_M k^2+V_S & 0 & -k \sqrt{1-\eta} V_M & 0 \\
 0 & V_M k^2+\frac{1}{V_S} & 0 & k \sqrt{1-\eta} V_M \\
 -k \sqrt{1-\eta} V_M & 0 & V_M+V_S-\eta (V_M+V_S-1) & 0 \\
 0 & k \sqrt{1-\eta} V_M & 0 & T+(1-\eta) \left(V_M+\frac{1}{V_S}\right) \\
\end{array}
\right),
\label{cm1}
\end{equation}

	where $\sigma_{B,LE}$ (\ref{ble}) describes correlations of Bob's mode $B$ with channel $E$ and leakage $L$ modes, $\gamma_{LE}$ is a covariance matrix of channel $E$ and leakage mode $L$. The conditional covariance matrix can be obtained as 

	\begin{equation}
\label{condition}
\gamma_{X|Y}=\gamma_{X}-\sigma_{Y,X}\left[\textbf{X} \gamma_Y \textbf{X} \right]^{MP} \sigma_{Y,X}^T, 
	\end{equation}
where $\textbf{X}=$ Diag(1,0,0,0) and $MP$ stands for Moore-Penrose pseudoinverse of the matrix \cite{Penrose1955}. Using Eq. (\ref{condition}) and elements of matrices (\ref{ble}) and (\ref{cm1}), as well as the matrix describing states received by Bob $\gamma_B=$Diag$([\eta(V_{S} +  V_{M}-1) + 1],0,0,[\eta(1/V_{S} +  V_{M}-1) + 1])$, one can find $V_{B|E}=V_{B|LE}=(V_M+k^2V_M+V_S)[\eta (k^2 V_M V_S^{-1}+1)+(1-\eta)(V_M + k^2 V_M +V_S)]^{-1}$. The latter can be used to assess Eve's mutual information with the trusted receiver side $I_{B E}= 1/2 \log_2 [V_B / V_{B|LE}]$, and consequently to find the key rate under individual attacks. 

    \vspace{\columnsep}
	\twocolumngrid

	\subsection{Purification} \label{Pure1}

	While P\&M schemes can be used for illustration of modus operandi of protocols and basic security analysis, for an extensive analysis of Gaussian CV QKD protocols one has to consider an entanglement-based scheme \cite{Reid}. The latter are based on usage of entangled sources that radiate two-mode Gaussian states described, in terms of covariance matrices, as
\begin{equation}
\label{CM}
\gamma=\left( 
\begin{matrix} V \mathbb{I} & \sqrt{V^2-1}\sigma_z \\ \sqrt{V^2-1}\sigma_z & V \mathbb{I} \end{matrix} \right),
\end{equation}
with $V$ being the variance of each mode, $\mathbb{I}$ is a two-dimensional unity matrix, and $\sigma_z$ is the Pauli matrix $\sigma_z=$Diag(1,0,0,-1). Alice performs a homodyne or heterodyne detection (depending on the protocol intended for use) on one of the modes, thereby conditionally squeezing the other mode, resulting in states with quadrature variances $1/V$ and $V$ or coherent states. The unmeasured and conditioned mode is a signal mode, that is sent through the untrusted quantum channel (characterized by losses $\eta$ and excess noise $\varepsilon$) to Bob. This technique yields fully equivalent results to P\&M schemes that prepare the signal state with the quadrature variance $V_S=1/V$ (or $V_S=1$ if Alice performs the heterodyne measurement), and subsequently apply Gaussian modulation of variance $V_M$.\par
The entanglement-based scheme with multimode modulation leakage should satisfy following conditions.
\begin{enumerate}
\item Neither states sent by Alice nor states received by Bob nor correlations $C_{AB}$ between them should be dependent on modulation ($kQ_{M}$, with variance $k^2V_M$) applied to states in the leakage mode.
\item The ratio between the leaking modulation and signal should be $k\geq 0$ and it's values can exceed 1, since generally the variance of the modulation applied can be greater than that applied to the signal mode. 
\item The ratio $k$ cannot be influenced by a trusted preparation party leaving only two parameters under Alice's control: signal modulation $Q_M$ and amount of squeezing in the state $Q_{S}$ produced by the source. 
\item The optical configuration should be scalable considering the fact that the trusted source can have an arbitrary multimodal structure.
\end{enumerate}

\begin{figure}[t]
\includegraphics[width=.75\linewidth]{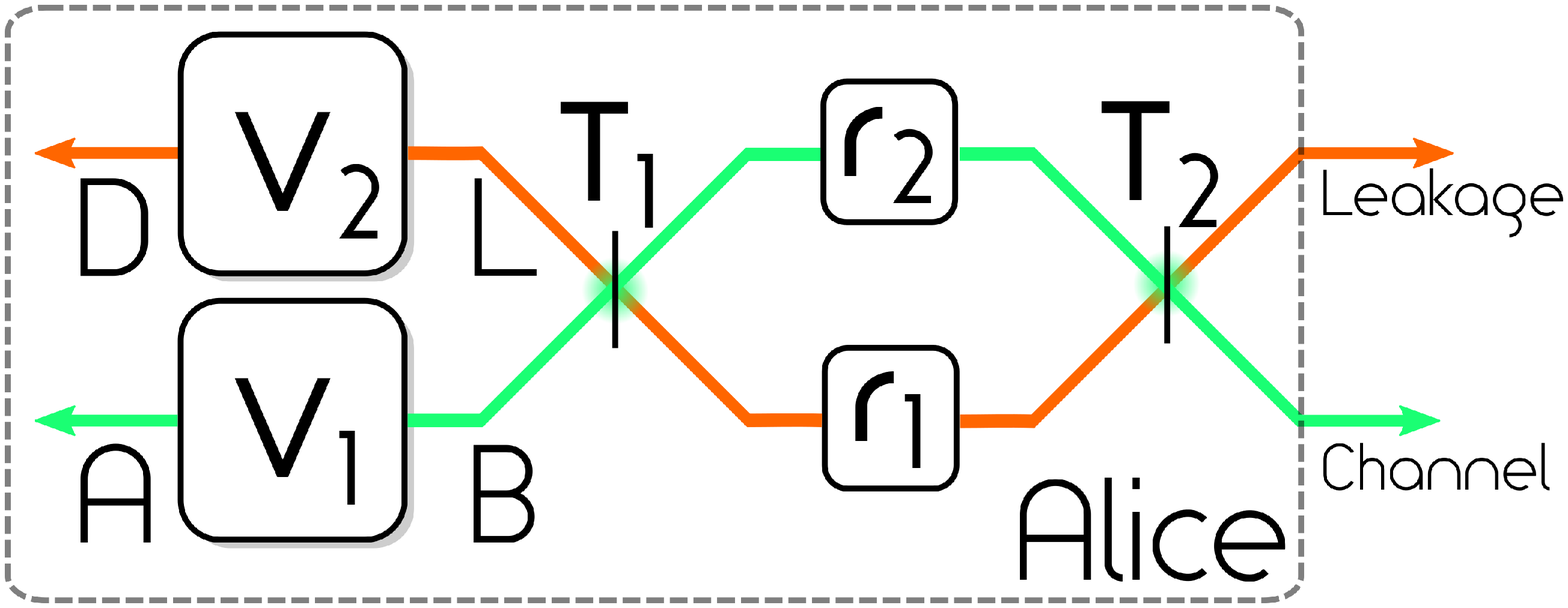}
	\caption{Purification of modulation leakage $(N=1)$ on the preparation side of the Gaussian CV QKD protocol. Alice's side contains two EPR sources, radiating modes $A$, $B$ with variance $V_1$, and modes $D$, $L$, with variance $V_2$. One mode from each source is kept on the preparation side ($A, D$) while the other two ($B, L$) interact on the beam splitter with transmittance $T_1$, undergo single-mode squeezing $r_1$ and $r_2$, respectively, and subsequently interact on the beam splitter with transmittance $T_2$. Signal mode $B$ proceeds through the untrusted channel ($\eta$, $\varepsilon$) to Bob, while the $L$ mode is accessible to Eve.}
	\label{EPRpure}
\end{figure}

    One of the solutions that can satisfy all required conditions is provided by the Bloch-Messiah decomposition theorem \cite{Braunstein2005a}, which says that the multimode evolution of an optical system governed by the linear Bogoliubov transformations can be decomposed into a combination of linear and nonlinear optical components (multiport interferometers, and single-mode squeezers). \par
	Let us consider the purification of two-mode modulation, i.e., the signal and leakage modes, as in Fig. \ref{EPRpure}. On the preparation side there are two sources; each generates a pair of entangled modes $A,B$ and $L,D$, respectively. The states $Q_A,Q_B$ ($Q$ represents $X$ or $P$ quadrature) in modes $A,B$ initially have a variance $V_1$, while states $Q_{L},Q_D$ in modes $L, D$ have variance $V_2$. One mode from each source, e.g., $B$ and $L$, interact on a beam splitter with transmittance $T_1$. The mode interaction effect on the covariance matrix $\gamma_{ABLD}$ is given by input-output relations (\ref{inout}).
	
	\begin{figure}[t]
\includegraphics[width=.75\linewidth]{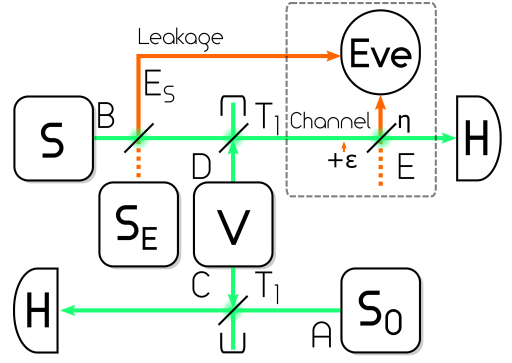}
	\caption{Purification of Gaussian CV QKD protocol with side channel between source and modulation. Source $S$ radiates signal (mode $B$) that, using entangled source $V$ (modes $C,D$), receives amplitude and phase modulation, and is sent to Bob, that conducts homodyne detection $H$. Source $S_0$ (mode $A$) generates infinitely squeezed states that are kept on preparation side. Source $S_E$ (mode $E_S$) establishes correlations with, and provides Eve with information about signal ($B$). Losses $\eta$ and noise $\varepsilon$ in untrusted channel (mode $E$) are attributed to Eve.} 
\label{PureS}	
\end{figure}

Further, states in modes $B$ and $L$ are squeezed on individual single-mode squeezers (characterized by the squeezing parameter $r_i$), resulting in the change of states quadrature variance by $e^{-2r_i}$ or $e^{2r_i}$. Subsequently modes interact, according to Eq. (\ref{inout}), on the beam splitter with transmittance $T_2$. As a result the four-mode covariance matrix $\gamma_{ABLD}$ after the interactions becomes $\gamma_{ABLD}'$ and depends on six parameters: $T_1, T_2, r_1, r_2, V_1, V_2$. The elements of the covariance matrix $\gamma_{ABLD}'$ can be used to form a set of equations:

\begin{equation}
    \label{set}
    \begin{aligned}
V_{B(X)} = &-2 t_1t_2e_{-} V_{-}+e^{-2 r_1} T_2 (T_1 V_{-}+V_2)\\
    &+e^{-2 r_2} (1-T_2) (V_1-T_1 V_{-}),\\
V_{B(P)} = & 2 t_1 t_2e_{-} V_{-}+e^{-2 r_1} (1-T_2) (T_1 V_{-}+V_2)\\
    &+e^{-2 r_2} T_2 (V_1-T_1 V_{-}),\\
V_{L(X)} = & -2 t_1t_2 e_{+} V_{-}+e^{2 r_1} T_2 (T_1 V_{-}+V_2)\\ 
    &+e^{2 r_2} (1-T_2) (V_1-T_1 V_{-}),\\
V_{L(P)} = & 2 t_1t_2 e_{+} V_{-}+e^{2 r_1} (1-T_2) (T_1 V_{-}+V_2)\\
    &+e^{2 r_2} T_2 (V_1-T_1 V_{-}),\\
C_{BL(X)} = & t_1(1-2 T_2)e_{-} V_{-}+t_2 (e^{-2 r_2} (V_1-T_1 V_{-})\\ 
    & -e^{-2 r_1} (T_1 V_{-}+V_2)),\\
C_{BL(P)} = & t_1(1-2 T_2) e_{+} V_{-}+t_2 (e^{2 r_2} (V_1-T_1 V_{-})\\ 
& -e^{2 r_1} (T_1V_{-}+V_2)).
    \end{aligned}
\end{equation}

where $V_{-}=V_1-V_2$, $t_{1(2)}=\sqrt{(1-T_{1(2)})T_{1(2)}}$, and $e_{\pm}=e^{\pm (r_1 + r_2)}$. To find the solutions of Eq. (\ref{set}) one can substitute the left-hand side for the respective variances of states in the signal, and leakage modes, and their covariances as follows: $V_{B(X)} \to V_S+V_M$, $V_{B(P)} \to 1/V_S+V_M$, $V_{L(X)}\to V_S+k^2V_M$, $V_{L(P)}\to 1/ V_S+k^2V_M$, and  $C_{BL(X)} \to k V_M$, $C_{BL(P)} \to - k V_M$. Solving Eq.(\ref{set}) for given $k, V_S, V_M$ will yield numerical values of parameters $T_1, T_2, r_1, r_2, V_1, V_2$ and subsequently a numerical covariance matrix $\gamma_{ABLD}'$ that can further be used to incorporate the effect of the untrusted quantum channel $(\eta, \varepsilon)$ and analyze the security of the Gaussian coherent- or squeezed-state CV QKD protocol. The same approach can be used to purify cases of $N$ mode leakage, however $N$ entangled sources are required, increasing the amount of parameters and equations in (\ref{set}) to $N(1+N)$. 

    \onecolumngrid
    \vspace{\columnsep}
\section{PREMODULATION LEAKAGE} \label{app2}
	\subsection{Pure losses}
		
Let us now consider the generic CV QKD protocol (without the multimode modulator) and the presence of the channel between the source and the modulator.  Results of linear interactions (\ref{inout}) in the P\&M scheme in purely lossy channel can be described by

\begin{equation}
\label{bee}
\sigma_{B,E_SE}=\left(
\begin{array}{cc}
(1-V_S)\sqrt{\eta_E \eta (1- \eta_E)} & 0 \\
 0 & -  \frac{1-V_S}{V_S}\sqrt{\eta_E \eta (1- \eta_E)} \\
 (\eta_E-V_M-\eta_E V_S)\sqrt{(1-\eta) \eta} & 0 \\
 0 & (\eta_E - \frac{\eta_E + V_M V_S}{V_S})\sqrt{(1-\eta) \eta} \\
\end{array}
\right),
\end{equation}

\begin{equation}
\label{ESE}
\sigma_{E_SE}=\left(
\begin{array}{cc}
(V_S-1) \sqrt{(1-\eta)(1-\eta_E) \eta_E} & 0\\
0 & \frac{(1-V_S)}{V_S} \sqrt{(1-\eta)(1-\eta_E) \eta_E} \\
\end{array}
\right),
\end{equation}

\begin{equation}
\label{E}
\gamma_{E}=\left(
\begin{array}{cc}
\eta +(1-\eta ) (V_M+\eta_E(V_S-1)+1) & 0 \\
0 & \left(V_M+\eta_E(V_S^{-1}-1)+1\right) (1-\eta )+\eta  \\
\end{array}
\right),
\end{equation}

\begin{equation}
\label{ES}
\gamma_{E_S}=\left(
\begin{array}{cc}
 V_S+(1-V_S) \eta_E & 0\\
0 & \frac{1+ \eta_E (V_S -1)}{V_S}\\
\end{array}
\right),
\end{equation}

where $\sigma_{B,E_SE}$ is the matrix describing Eve's correlations to the signal mode after premodulation leakage and losses, $\sigma_{E_SE}$ describes correlations between modes accessible to Eve, and the variances (in $X$ and $P$ quadratures) of the latter are given by $\gamma_{E}$, and $\gamma_{E_S}$. One can use Eqs. (\ref{condition}) and (\ref{bee})--(\ref{ES}) to find the variance of Alice and Bob states conditioned by measurements of modes accessible to Eve:

\begin{equation}
V_{A|E}= \frac{V_S[V_M(1 + (1-\eta)V_S) + V_S]+ \eta_E(1-V_S)[\eta(V_M-V_S)+(1-\eta) V_M V_S]}{V_S[1-\eta+(1-\eta)V_M + \eta] + \eta_E(1-V_S)[(1-\eta)V_M + \eta]},
\label{VAE}
\end{equation}

\begin{equation}
V_{B|E}= \frac{\eta_E V_M + V_S(V_M(1-\eta_E)+1)}{V_S[1+\eta_E(1-V_S)((1-\eta)V_M + \eta)+(1-\eta) V_M]},
\label{VBE}
\end{equation}

Eve's mutual information with a trusted party $I_{A E}=1/2 \log_2 [V_A/V_{A|E}]$, or $I_{BE}=1/2 \log_2 [V_B/V_{B|E}]$ can be calculated using, respectively, Eq. (\ref{VAE}) or (\ref{VBE}) and further used to assess the key rate under individual attacks. 

	\subsection{Purification}

In the case of the side channel present between the source and the modulator, the purification can similarly be done using Bloch-Messiah decomposition (as in Appendix \ref{Pure1}), however, we adopt a general purification scheme as in Fig. \ref{PureS} \cite{sidech1}. \\ Alice on the preparation side operates an EPR source (\ref{CM}) that radiates into modes $C$ and $D$, source $S$ that produces the signal state in mode $B$ and source $S_0$ that produces the infinitely squeezed state in mode $A$. Modes produced by the EPR source have variance $V_M/(1-T_1)$ and are, respectively, coupled to modes from other the two sources on strongly unbalanced beam splitters $T_1$. The leakage is modelled as a signal interaction (\ref{inout}) with the vacuum mode on a beam splitter with transmittance $\eta_E$. The signal further proceeds to the unbalanced beam splitter $T_1$ where it interacts with mode $D$ that carries information and further is sent to the untrusted channel where it suffers from losses $\eta$ and noise $\varepsilon$. Mode $A$ carrying the infinitely squeezed state (to simulate the detection on the trusted side) interacts with first entangled mode $C$ on another strongly unbalanced beam splitter characterized by the same value of transmittance $T_1$. \\ After the interactions, the state that is kept on the preparation side can be described by the variances as: $V_{A(X)}=V_{S_0}T_1+V_M$ ,  $V_{A(P)}=T_1/(V_{S_0})+V_M$, and the state that is sent to Bob through the untrusted channel as: $V_B=(V_S \eta_E + (1-\eta_E)V_{E_1})T_1+V_M$, while these two states are correlated as $C_{AB}=-\sqrt{(V^2_M-(1-T_1)^2)\eta}$. In the limit $T_1 \to1$ these correspond to the P\&M scheme with the premodulation channel---generation of the signal state with variance $V_S$, the premodulation channel interaction with output accessible to Eve and further amplitude and/or phase Gaussian modulation of variance $V_M$. The measurement conducted by Alice conditions Bob's state to: $V_{B|A (X)}=T_1 ((1-\eta_E)V_{E_1}+\eta_E V_S)+\frac{(T_1-1)^2-V_M^2}{T_1 V_0+V_M}+V_M$, $V_{B|A (P)}=T_1 \left((1-\eta_E) V_{E_1}+\frac{\eta_E}{V_S}\right)+\frac{V_0 \left((T_1-1)^2-V_M^2\right)}{T_1+V_0 V_M}+V_M$. In the regime $T_1 = 1$, $V_0 = 1$ and $V_{E_1}=1$, $\eta_E=1$ states reduce to $V_{B|A (X)}=V_S$, and $V_{B|A (P)}=1/V_S+V_M$ that corresponds to modulation with variance $V_M$ applied to both quadratures of the signal state described initially by variances in respective quadratures $V_S$, $1/V_S$ with only one value $(x)$ being kept. The resulting six-mode covariance matrix (including premodulation $E_S$ and untrusted $E$ channels) $\gamma_{ABCDE_SE}$ allows one to further analyze the security of the Gaussian coherent- or squeezed-state CV QKD protocol.

\twocolumngrid

\end{document}